\documentclass[conference]{IEEEtran}

\usepackage{cite}
\usepackage{amsmath,amssymb,amsfonts}
\usepackage{algorithmic}
\usepackage{graphicx}
\usepackage{textcomp}
\usepackage{physics}
\usepackage{hyperref}
\usepackage{xcolor}
\usepackage{float}
\def\BibTeX{{\rm B\kern-.05em{\sc i\kern-.025em b}\kern-.08em
    T\kern-.1667em\lower.7ex\hbox{E}\kern-.125emX}}
\begin{document}

\title{Amplification based on the noise-induced negative differential resistance in a Zener diode}

\author{\IEEEauthorblockN{Alexandre Dumont}
\IEEEauthorblockA{
\textit{Institut Quantique,}\\
\textit{Département de physique,} \\
\textit{Université de Sherbrooke}\\
Sherbrooke, Canada \\
Alexandre.Dumont3@usherbrooke.ca}
\and
\IEEEauthorblockN{Bertrand Reulet}
\IEEEauthorblockA{
\textit{Institut Quantique,}\\
\textit{Département de physique,} \\
\textit{Université de Sherbrooke}\\
Sherbrooke, Canada \\
Bertrand.Reulet@usherbrooke.ca}
}

\maketitle

\begin{abstract}
A voltage biased Zener diode always exhibit positive differential resistance, thus cannot 
be used as an element to provide amplification of a signal. 
We show how to induce negative differential resistance in the reverse bias regime of 
a 12V Zener diode by noise feedback. 
We use this to build a voltage amplifier in the audio frequency range, which we 
characterize by providing bandwidth, gain, power consumption, gain compression and output noise spectral density.
\end{abstract}


\section{Introduction}
It is well accepted that one cannot amplify a signal using only passive components such 
as diodes, resistors, capacitors and inductors together with a voltage source. 
We demonstrate here that on the contrary, one can. 
This is made possible by using the effects of feedback of the noise generated by the diode 
in the circuit in which it is embedded: while a Zener diode taken alone does not exhibit 
negative differential resistance, the same diode in a well engineered passive, linear 
circuit may\cite{thibault_noise_2021}.

There are amplifiers based on specific kinds of diodes, in particular tunnel and IMPATT diodes. 
The tunnel diode when forward biased has an I-V characteristic that exhibits a region of 
negative differential resistance, when quantum tunneling occurs~\cite{esaki_new_1958}. 
This can be used to amplify a signal~\cite{miles_microwave_1965}. 
Similarly, an IMPATT diode exhibits negative differential resistance because of finite 
transit time ~\cite{shockley_negative_1954}, which allows to build amplifiers working in 
the microwave domain~\cite{lee_50-ghz_1968}. 
Here we use a Zener diode which does not have negative differential resistance, but is 
very noisy. It is the fact that the noise has a strong bias dependence, together with 
the use of a large series resistor, that makes amplification possible. 

This paper is structured as follows: in section II, we present the 
requirements for feedback effects to affect a device I-V characteristics from a 
theoretical standpoint. In section III, we show how a resistor in series with the diode 
modifies its I-V characteristics, which may exhibit negative differential resistance. 
Section IV focuses on the design and considerations going into making the Zener 
diode and its immediate environment into the heart of a low frequency voltage amplifier.

\section{Theory of noise feedback}~\label{sec:theory}
This work is based on the noise feedback mechanism,  
demonstrated theoretically and experimentally in~\cite{thibault_noise_2021}. This mechanism makes  
 it incorrect to assume that a component has an intrinsic DC I-V 
characteristics, independent of the circuit the device is connected to. 
On the contrary, the electromagnetic environment of the device, i.e. the circuit it is 
connected to, modifies the I-V. 
This occurs because the component interacts with its own noise through 
the external impedance connected to it. 
The effect is larger when the noise generated by the component has a strong voltage 
dependence, as reflected by the following equation:
\begin{align}
    \pdv{V(I,R)}{R} &= \frac{1}{2}\pdv{S_{I}(I,R)}{I},
\end{align}
where $V(I,R)$ is the average, dc voltage across the device, $I$ 
the average DC current through the device and $S_{I}(I,R)$ is the variance of 
the current fluctuations generated by the device. 
Both $V$ and $S_I$ depend on the environmental resistance $R$ in series with 
the device.  
Here we consider a Zener diode, which noise is strongly voltage dependent close to the 
breakdown voltage, to the point that it may lead to the diode exhibiting a negative 
differential resistance when in series with $R$. 
Note that such a property could occur with potentially any component that exhibits 
a strong enough bias-dependent noise. 
For a small resistance $R$, the correction to the IV characteristic is given by:
\begin{equation}
    \Delta V(I)=\frac R2 \frac{\partial S_{I}}{\partial I}
    \label{eq:dV}
\end{equation}
A region in bias with negative differential resistance may thus occur for a large 
enough $R$ and for a current noise that is a concave function of the current, 
$\frac{\partial^2 S_{I}}{\partial I^2}<0$. This is precisely what we observe, 
see Fig.\ref{fig:Diode_dV_dI}. 

\section{Effects of feedback on a 1N759A Zener diode}
\begin{figure}[H]
	\centering
	\includegraphics[width=\columnwidth]{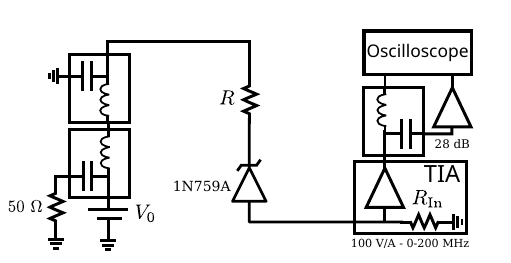}
	\caption{Measurement setup for the I-V characteristics of the diode. TIA represents a trans-impedance amplifier, FEMTO model DHCPA-100. Bias tees are Mini-Circuits ZFBT-6GWB+.}
	\label{fig:Setup_IV}
\end{figure}

Since the I-V characteristics depends on the rest of the circuit, all the measurements need 
to be performed with the same circuit where the impedance seen by the diode is well 
controlled at all relevant frequencies.  
To this end, we use the setup shown in Fig.~\ref{fig:Setup_IV}. 
Two identical bias-tees are used to connect the voltage source $V_0$ to the diode and its series 
resistor, the first one redirects the noise of the voltage source to a 50~$\Omega$ resistor 
and the second one provides a known RF impedance to essentially shunt the inductance from 
the diode's point-of-view.

The current is measured by a trans-impedance amplifier (TIA) with a $R_\mathrm{In}=50~\Omega$ resistor 
input impedance. 
The voltage at the output of the TIA is split into two frequency bands by a bias tee. 
The low frequency part is used to measure the dc voltage, proportional to the dc current 
in the circuit. 
The high frequency part is used to characterize the current noise in the circuit. 
Separating the two allows to further amplify the current fluctuations before digitization.

\begin{figure}[H]
	\centering
	\includegraphics[width=\columnwidth]{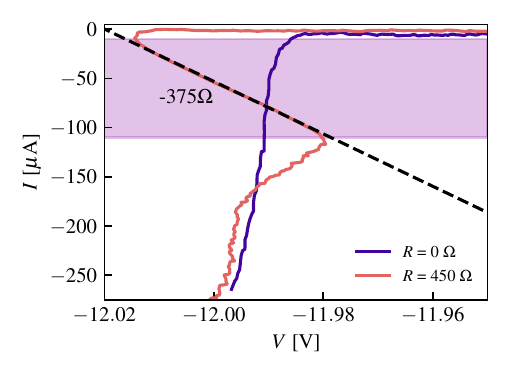}
	\caption{$I(V)$ characteristics of the diode for two values of environmental resistance. 
    The highlighted section is the negative differential resistance region.}
	\label{fig:Diode_IV}
\end{figure}

\begin{figure}[H]
	\centering
	\includegraphics[width=\columnwidth]{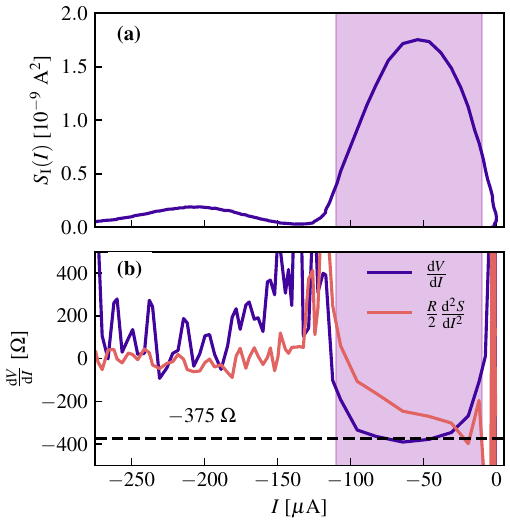}
	\caption{\textbf{(a)} Variance of the current fluctuations at $R=450$~$\Omega$ integrated from 
    100 kHz to 200 MHz
    \textbf{(b)} Differential resistance of the diode, and normalized second derivative of its integrated noise. Both are taken at $R=450$~$\Omega$.}
	\label{fig:Diode_dV_dI}
\end{figure}

Fig.~\ref{fig:Diode_IV} shows the $I(V)$ curves of the diode for $R=0~\Omega$ 
(the total resistance in the circuit is the  input impedance of the TIA, $R_\mathrm{In}=50~\Omega$) 
and $R=450~\Omega$. 
The voltage $V$ across the diode is obtained by $V=V_0-I/(R+R_\mathrm{In})$. 
The curve obtained for $R=0$ is the usual one for a Zener diode with a threshold voltage 
close to 12V. 
For $R=450~\Omega$, the result is strikingly different: the threshold is 
shifted by about 30~mV and instead of the familiar sharp increase of the (negative) current 
we observe that the current is a non-monotonous function of the voltage: there is a region 
of negative slope, i.e. negative differential resistance, indicated by a shaded area. 
The dashed black line corresponds to a resistance of $-375~\Omega$.
This anomaly appears only close to the threshold. 
The differential resistance $dV/dI$ is shown as a function of the dc current $I$ in 
Fig.\ref{fig:Diode_dV_dI}(b) 

Fig.~\ref{fig:Diode_dV_dI}(a) shows the variance of the current fluctuations 
integrated over frequency between 100 kHz to 200 MHz~\cite{dumont_thermodynamique_2025}. 
As discussed above, there is a region where the noise is a concave function 
of the current (shaded area). 
It corresponds to the region of negative differential resistance $dV/dI\sim-375~\Omega$ 
between $\sim-25~\mu$A and $\sim-100~\mu$A.
This determines the biasing condition of our amplifier. 
In this region the noise is huge, with a standard deviation of $\sim40~\mu$A at the maximum. 
The theoretical prediction of Eq.(\ref{eq:dV}) is plotted in red. 
It is close to the measured value of $dV/dI$.
The power dissipated in the diode is typically 
$P\simeq12\mathrm{V}\times 50\mu\mathrm{A}\simeq0.6$~mW.
At higher current the differential resistance is very noisy but fluctuates 
around $\sim100~\Omega$. 
There is another region of negative concavity of $S_I(I)$. 
However, the noise there is not large enough to provide negative differential resistance.

\section{Amplifier design}
\begin{figure}[H]
	\centering
	\includegraphics[width=\columnwidth]{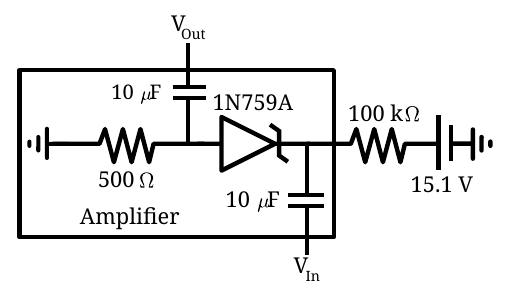}
	\caption{Electrical schematics of the Zener amplifier.}
	\label{fig:Amplifier_Schematics}
\end{figure}

The design of the amplifier is very simple, relying on a voltage divider, 
since the goal is not to create a state-of-the-art amplifier but rather to 
have a proof of concept. 
The voltage divider involves the (negative) differential resistance $r$ of the diode and 
a resistor $R$: 
\begin{align}
    G=\frac{V_{\rm Out}}{V_{\rm In}} = \frac{R}{R+r}.
\end{align}
For $-|R|<r<0$, $G>1$: there is voltage amplification. 
Fig.~\ref{fig:Amplifier_Schematics} shows the amplifier schematics. 
The amplifier is current biased by a 15.1V voltage source in series with a 
$100$~k$\Omega$ resistor. 
A fixed current bias is easy to achieve because the $I(V)$ curve of the diode is single 
valued. In contrast, the $V(I)$ is multivalued, making a voltage bias more difficult. 
Since there is a $12$V voltage drop across the diode, the biasing current is 
$I_{\rm bias}=31~\mu\rm A$. This biasing condition corresponds to the optimum voltage gain. 
The total power dissipated by the amplifier is $P=0.47~\rm mW$, which 
includes $0.1~\rm mW$ dissipated in the 100~k$\Omega$ resistor.   
The input signal is fed into the amplifier via a $C_{\rm In }=10~\mu$F coupling capacitor 
connected to the front of the diode while the output is taken in between the diode and the 
resistor $R=500~\Omega$ to create the voltage divider. 

\section{Amplifier characterization}
\begin{figure}[H]
	\centering
	\includegraphics[width=\columnwidth]{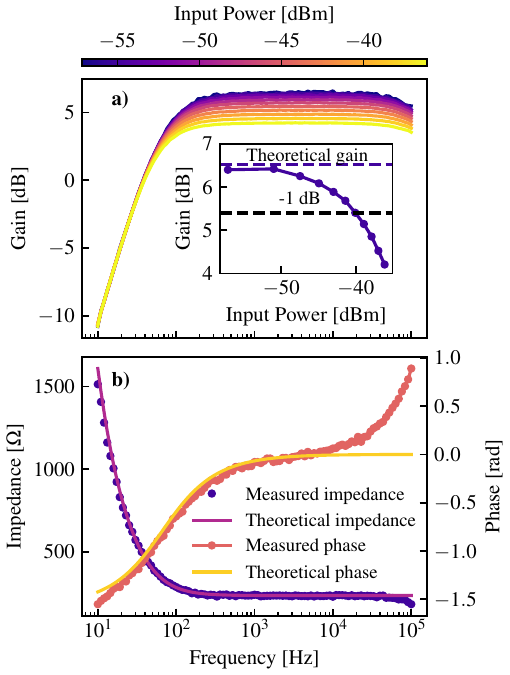}
	\caption{(a) Amplifier voltage gain as a function of frequency for various input powers. Inset: gain vs. input power at 1 kHz. 
    (b) Modulus and phase of the input impedance as a function of frequency.}
	\label{fig:Gain_compression}
\end{figure}

We now characterize the amplifier by measuring its input impedance, gain, bandwidth, compression and noise. 
To do so we connect a source of sine wave (Agilent 33522A) of varying frequency and amplitude at 
the input of the voltage and measure the input $V_{\rm In}$ at output $V_{\rm Out}$ voltages 
with two Stanford Research model SR830 lockin amplifiers synchronized with the source. 
The input impedance of the amplifier is given by $Z_{\rm In}=R_0V_{\rm In}/(V_0-V_{\rm In})$ 
with $R_0=50~\Omega$ the output impedance of the source, and $V_0$ the voltage it applies, 
see Fig.\ref{fig:Gain_compression}(b). 
We find $Z_{\rm In}\simeq236~\Omega$ (real) whithin the bandwidth of the amplifier. 
Theory predicts $Z_{\rm In}=R+r+1/(jC_{\rm In }\omega)$ from which we deduce $r=-264$~$\Omega$. 
This value is not as negative as what has been measured with the previous setup. 
The exact value of $r$ depends on the impedance seen by the diode over a very large 
bandwidth, that where it produces noise, i.e. $\sim200~\rm MHz$. 
This frequency-dependent environmental impedance differs in both circuits. 

The voltage gain in dB, $G_{\rm dB}=20\log(V_{\rm Out}/V_{\rm In})$ is shown 
in Fig.~\ref{fig:Gain_compression}(a) as a function of frequency between 10~Hz and 100~kHz, 
for various input powers between -50~dBm and -36~dBm. 
At low input power the gain is independent of the input power and reaches $\sim6.5~\rm dB$, 
in very good agreement with the theoretical expectation $R/(R+r)\simeq6.74~\rm dB$ using 
the measured value of $r$. 
At high power the amplifier saturates and the gain decreases, as usual. 
We show in the inset of Fig.~\ref{fig:Gain_compression}(a) the gain measured at 1~kHz vs. 
input power. 
We deduce a 1~dB voltage compression point of -40~dBm. 
The theoretical expectation for the gain is $G=R/|Z_{\rm In}|$, shown as a purple dashed 
line in the inset of Fig.\ref{fig:Gain_compression}(a), taking the measured value for $Z_{\rm In}$. 
At high frequency we observe a slight decrease of the gain and input impedance, which 
acquires an inductive component.    

Considering now the frequency dependence of the gain, we observe a low frequency cutoff at 70~Hz. 
This value is well explained by the input capacitor and the input impedance model. 
The gain then  rises steadily to reach a plateau at around 6.5~dB until it 
starts declining again at around 50~kHz. 
This decline is not yet explained, but this still allows for a bandwidth spanning from 
70~Hz to 100~kHz when considering the 3~dB cutoff frequencies. 
The gain of 6.5~dB is also very close to the theoretical maximum gain of 6.74~dB which 
indicates that the estimate of the negative differential resistance from the input 
impedance is correct.
The high frequency cutoff above 100~kHz is accompanied by a sharp increase in 
phase. 
Its origin is not yet understood. 
It does not come from the input impedance which remains almost flat. 
It most probably comes from the fact that the noise generated by the diode does not 
react instantaneously to the bias voltage, in other words the noise susceptibility, i.e. 
the linear response of noise to a varying voltage, no longer reduces to $dS_I/dV$ but 
acquires a frequency dependence \cite{gabelli_noise_2007}. 
These effects should be taken into account in the theory, and their measurements require 
subtle studies of noise correlations induced by the ac voltage \cite{farley_sensing_2025}.

\begin{figure}[H]
	\centering
	\includegraphics[width=\columnwidth]{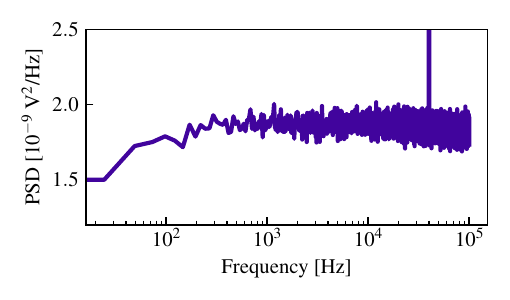}
	\caption{Power spectral density at the output of the amplifier.}
	\label{fig:PSD}
\end{figure}

The voltage noise power spectral density of the amplifier is shown in Fig.~\ref{fig:PSD}. 
As expected, the amplifier is very noisy. 
To get rid of that noise, it would be necessary to separate the signals into two bands: 
one with low noise, where the signal of interest is amplified, and a separate one where the 
noisy component generates its noise to give rise to the noise feedback effect. 
With a Zener diode, both bands cannot be separated, but other components may behave 
differently.   

The spike at 40 kHz is parasitic and not due to the amplifier.

\section{Conclusions}
We report on a Zener diode based amplifier which, against usual predictions, 
is made possible due to the effects of noise feedback between the diode and a 
resistive environment. 
The amplifier is characterized by a bandwidth spanning from 70~Hz to 100~kHz with 
a nominal gain of about 6.44~dB and a 1~dB compression point at -40~dBm of input 
power. 
The phenomenon we have demonstrated is very generic. It applies in principle to any circuit. 
Our work paves the way towards the use of noise feedback in other kinds of circuits 
using components which noise strongly depends on bias.

\section*{Acknowledgements}
We thank Karl Thibault for fruitful discussion and Christian Lupien for technical help. 
This work has been supported by the Canada Research Chair program, the NSERC, the Canada 
First Research Excellence Fund, the FRQNT, and the Canada Foundation for Innovation.

\bibliographystyle{ieeetr}
\bibliography{bibliography}

@article{shockley_negative_1954,
	title = {Negative resistance arising from transit time in semiconductor diodes},
	volume = {33},
	issn = {0005-8580},
	url = {https://ieeexplore.ieee.org/document/6768393},
	doi = {10.1002/j.1538-7305.1954.tb03742.x},
	number = {4},
	urldate = {2024-10-11},
	journal = {The Bell System Technical Journal},
	author = {Shockley, W.},
	month = jul,
	year = {1954},
	note = {Conference Name: The Bell System Technical Journal},
	pages = {799--826},
}

@article{esaki_new_1958,
	title = {New {Phenomenon} in {Narrow} {Germanium} p-n {Junctions}},
	volume = {109},
	url = {https://link.aps.org/doi/10.1103/PhysRev.109.603},
	doi = {10.1103/PhysRev.109.603},
	number = {2},
	urldate = {2024-10-13},
	journal = {Physical Review},
	author = {Esaki, Leo},
	month = jan,
	year = {1958},
	note = {Publisher: American Physical Society},
	pages = {603--604},
}

@article{thibault_noise_2021,
	title = {Noise feedback in an electronic circuit},
	volume = {3},
	url = {https://link.aps.org/doi/10.1103/PhysRevResearch.3.033058},
	doi = {10.1103/PhysRevResearch.3.033058},
	number = {3},
	urldate = {2024-10-18},
	journal = {Physical Review Research},
	author = {Thibault, Karl and Gabelli, Julien and Lupien, Christian and Reulet, Bertrand},
	month = jul,
	year = {2021},
	note = {Publisher: American Physical Society},
	pages = {033058},
}

@article{lee_50-ghz_1968,
	title = {A 50-{GHz} silicon {IMPATT} diode oscillator and amplifier},
	volume = {15},
	issn = {1557-9646},
	url = {https://ieeexplore.ieee.org/abstract/document/1475410},
	doi = {10.1109/T-ED.1968.16508},
	number = {10},
	urldate = {2025-04-02},
	journal = {IEEE Transactions on Electron Devices},
	author = {Lee, T.P. and Standley, R.D. and Misawa, T.},
	month = oct,
	year = {1968},
	note = {Conference Name: IEEE Transactions on Electron Devices},
	pages = {741--747},
}

@inproceedings{miles_microwave_1965,
	title = {A microwave tunnel-diode amplifier in stripline},
	volume = {VIII},
	url = {https://ieeexplore.ieee.org/abstract/document/1157583/references#references},
	doi = {10.1109/ISSCC.1965.1157583},
	urldate = {2025-04-02},
	booktitle = {1965 {IEEE} {International} {Solid}-{State} {Circuits} {Conference}. {Digest} of {Technical} {Papers}},
	author = {Miles, T. and Cox, D.},
	month = feb,
	year = {1965},
	pages = {24--25},
}

@article{farley_sensing_2025,
	title = {Sensing {Quantum} {Vacuum} {Fluctuations} with {Non}-{Gaussian} {Electronic} {Noise}},
	volume = {134},
	url = {https://link.aps.org/doi/10.1103/PhysRevLett.134.166301},
	doi = {10.1103/PhysRevLett.134.166301},
	number = {16},
	urldate = {2025-12-17},
	journal = {Physical Review Letters},
	author = {Farley, Clovis and Pinsolle, Edouard and Reulet, Bertrand},
	month = apr,
	year = {2025},
	note = {Publisher: American Physical Society},
	pages = {166301},
}

@inproceedings{gabelli_noise_2007,
	title = {The noise susceptibility of a coherent conductor},
	volume = {6600},
	url = {https://www.spiedigitallibrary.org/conference-proceedings-of-spie/6600/66000T/The-noise-susceptibility-of-a-coherent-conductor/10.1117/12.724656.full},
	doi = {10.1117/12.724656},
	urldate = {2025-12-17},
	booktitle = {Noise and {Fluctuations} in {Circuits}, {Devices}, and {Materials}},
	publisher = {SPIE},
	author = {Gabelli, J. and Reulet, B.},
	month = jun,
	year = {2007},
	pages = {246--257},
}

@phdthesis{dumont_thermodynamique_2025,
	title = {Thermodynamique de {Fourier} dans les circuits micro-ondes et effets de rétroaction du bruit dans la statistique des fluctuations de courant d'une composante électronique},
	url = {https://hdl.handle.net/11143/23208},
	language = {fr},
	urldate = {2025-12-17},
	school = {Université de Sherbrooke},
	author = {Dumont, Alexandre},
	year = {2025},
}

\end{document}